\documentclass[onecolumn,showpacs,amsmath,amssymb]{revtex4}

\usepackage{amsmath,color}
\usepackage{graphicx}
\usepackage{dcolumn}
\usepackage{bm}

\begin{document}

\preprint{APS/123-QED}

\title{Extraordinary surface voltage effect in the invisibility cloak with an active device inside}

\author{Baile Zhang}
\author{Hongsheng Chen}
\email{chenhs@ewt.mit.edu}
\author{Bae-Ian Wu}
\author{Jin Au Kong}%
\affiliation{The Electromagnetics Academy at Zhejiang University, Zhejiang University, Hangzhou 310058, China, and Research Laboratory of Electronics, Massachusetts Institute of Technology, Cambridge, MA 02139, USA.}

\begin{abstract}
The electromagnetic field solution for a spherical invisibility
cloak with an active device inside is established. Extraordinary
electric and magnetic surface voltages are induced at the inner
boundary of a spherical cloak, which prevent electromagnetic waves
from going out. The phase and handness of polarized waves obliquely
incident on such boundaries is kept in the reflected waves. The
surface voltages due to an electric dipole inside the concealed
region are found equal to the auxiliary scalar potentials at the
inner boundary, which consequently gain physical counterparts in
this case.
\end{abstract}

\maketitle

Pendry \emph{et al.} \cite{pendry} firstly proposed the invisibility
cloak based on coordinate transforms where a ``hole'' is created in
the transformed coordinate system and an object in the ``hole'' can
be concealed from detection. Another method was reported to produce
similar effects in the geometric limit \cite{leonhardt}. Designs of
invisibility cloaks based on other methods have also been published
\cite{greenleaf_static,alu,milton,sihvola}. From the viewpoint of
transformation method \cite{pendry}, the hole creation does not
result in an electromagnetic vacuum but rather a complete separation
of electromagnetic domains into a cloaked region and those outside
\cite{pendry,sheng}.  More precisely, a ``true'' cloak should not
only cloak passive objects from incoming waves, but also cloak
active devices by preventing waves from going out and being
detected. The effectiveness of a transformation based cloak design
for hiding a passive object was first confirmed computationally in
the geometric optics limit \cite{pendry,schurig_ray}, in full-wave
finite-element simulations \cite{cummer,zolla}, and in full-wave
analytic scattering models with deeper physical interpretations
\cite{hongsheng,baile}. It has also been demonstrated experimentally
with a simplified practical model \cite{schurig}. On the other hand,
however, the electromagnetic wave behavior in this concealed region
with an active device inside remains unknown, due to the fact that
the concealed region, or the ``hole'', created by the transformation
method does not exist before transformation and has no counterpart
in the original Euclidian space. A rigorous mathematical treatment
of cloaking shows that ``finite energy solutions'' to Maxwell's
equations do not exist in the presence of active sources inside the
concealed region \cite{greenleaf}, which strongly implies that some
special physical phenomenon must exist but has not been revealed.
Also, the exact electromagnetic field solution to the Maxwell
equations in this case, as well as whether waves can go out of the
concealed region, still haven't been established or confirmed.

In this Letter, the exact field solution to the Maxwell equations
with an active source inside the concealed region of a spherical
cloak is established. We show that, electric and magnetic surface
voltages are induced due to an infinite polarization of the material
at the inner boundary, which prevent the electromagnetic waves from
going out. The special property of the material at the inner
boundary of the cloak, is able to keep not only the handness of
polarization, but also the phase information, in the waves reflected
from the inner boundary, whose behavior is different from a perfect
electric/magnetic conductor (PEC/PMC), or the artificial soft and
hard surface (SHS) in electrical engineering \cite{lindell}. The
induced surface electric and magnetic voltages are shown to be
exactly equal to the auxiliary scalar electric and magnetic
potentials at the inner boundary, respectively. Therefore, these
auxiliary potentials, which are commonly introduced in accompany
with vector potentials as strictly mathematical tools for most
engineers \cite{balanis}, have physical counterparts at the inner
boundary in this case.

\begin{figure}
\includegraphics[width=0.25\columnwidth,draft=false]{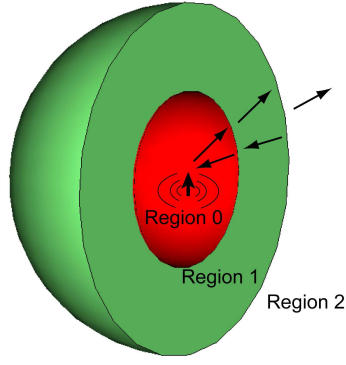}
\caption{\label{fig:sphere}(Color online) Reflection  and
transmission of electromagnetic waves with an electric dipole
inside the concealed region of a spherical cloak. Region 0:
concealed region; Region 1: cloak layer; Region 2: free space.}
\end{figure}

Figure~\ref{fig:sphere} shows the configuration of  a spherical
cloak with outer radius $R_2$ and inner radius $R_1$. The cloak
layer within $R_1<r<R_2$ is a specified anisotropic and
inhomogeneous medium with permittivity tensor
$\bar{\bar{\epsilon}}=\epsilon_r
\hat{r}\hat{r}+\epsilon_t\hat{\theta}\hat{\theta}+\epsilon_t\hat{\phi}\hat{\phi}$
and permeability tensor $\bar{\bar{\mu}}=\mu_r
\hat{r}\hat{r}+\mu_t\hat{\theta}\hat{\theta}+\mu_t\hat{\phi}\hat{\phi}$.
According to Ref. \cite{pendry}, it is chosen such that
$\epsilon_t/\epsilon_0=\mu_t/\mu_0=R_2/(R_2-R_1)$ and
$\epsilon_r/\epsilon_t=\mu_r/\mu_t=(r-R_1)^2/r^2$. Without loss of
generality, we assume that the background material in the region
$r<R_1$ has permittivity $\epsilon_1$ and permeability $\mu_1$. A
time-harmonic electric dipole is put inside as an active device. The
electromagnetic waves from the dipole as well as the response of the
surrounding environment can be decomposed into TE and TM modes with
respect to $\hat{r}$, corresponding to scalar potentials $\Psi_{TE}$
and $\Psi_{TM}$, whose expressions in the current case have been
derived in Ref. \cite{hongsheng}.

Since TE and TM modes in such a radially inhomogeneous medium can be
shown to be decoupled \cite{weng}, the derivations for these two
kinds of modes
 are identical to each other. We start with the case that an
outgoing TM wave is excited in the concealed region. This outgoing
wave will induce a standing wave in region $r<R_1$, an outgoing wave
and a standing wave in region $R_1<r<R_2$, and an outgoing wave in
region $r>R_2$ \cite{weng}, as shown in Fig.~\ref{fig:sphere}. Thus
the scalar potentials in these three regions can be written as
\begin{eqnarray}
\Psi_{TM}^{int}&=&[\zeta_n(k_1r)+R^{TM}\psi_n(k_1r)]P_n^m(\cos
\theta)e^{im\phi}\;
\\
\Psi_{TM}^{c}&=&[d_n^M\psi_n(k_t(r-R_1))+f_n^M\chi_n(k_t(r-R_1))]\nonumber
\\
&\cdot&P_n^m(\cos \theta)e^{im\phi}\;
\\
\Psi_{TM}^{out}&=&T^{TM}\zeta_n(k_0r)P_n^m(\cos \theta)e^{im\phi}\;
\end{eqnarray}
where $\psi_n$, $\chi_n$, and $\zeta_n$ are Riccati-Bessel Functions
of the first, the second, and the third kind respectively; $R^{TM}$,
$d_n^M$, $f_n^M$, and $T^{TM}$ are the unknown expansion
coefficients. Especially, $R^{TM}$ and $T^{TM}$ are called the
general reflection coefficient and general transmission coefficient,
respectively \cite{weng}.

For the sake of illustration, the inner boundary of the cloak is set
at $r=R_1+\delta$ instead of $r=R_1$ and then the limit $\delta
\rightarrow 0$ is taken \cite{baile}. Consequently, four boundary
equations can be listed utilizing the continuities of tangential $E$
and tangential $H$ at the outer boundary (Eq.(\ref{eq:E_R2}) and
(\ref{eq:H_R2})) and at the inner boundary (Eq.(\ref{eq:E_R1}) and
(\ref{eq:H_R1})):
\begin{eqnarray}
&1/\sqrt{\mu_t
\epsilon_t}[d_n^M\psi_n^{'}(k_t(R_2-R_1))+f_n^M\chi_n^{'}(k_t(R_2-R_1))]&\nonumber
\\
&=1/\sqrt{\mu_0\epsilon_0}T^{TM}\zeta_n^{'}(k_0R_2)\;&\label{eq:E_R2}
\\
&1/\mu_t[d_n^M\psi_n(k_t(R_2-R_1))+f_n^M\chi_n(k_t(R_2-R_1))]&\nonumber
\\
&=1/\mu_0T^{TM}\zeta_n(k_0R_2)\;&\label{eq:H_R2}
\\
&1/\sqrt{\mu_1
\epsilon_1}[\zeta_n^{'}(k_1(R_1+\delta))+R^{TM}\psi_n^{'}(k_1(R_1+\delta))]&\nonumber
\\
&=1/\sqrt{\mu_t\epsilon_t}[d_n^M\psi_n^{'}(k_t\delta)+f_n^M\chi_n^{'}(k_t\delta)]\;&\label{eq:E_R1}
\\
&1/\mu_1[\zeta_n(k_1(R_1+\delta))+R^{TM}\psi_n(k_1(R_1+\delta))]&\nonumber
\\
&=1/\mu_t[d_n^M\psi_n(k_t\delta)+f_n^M\chi_n(k_t\delta)]\;&\label{eq:H_R1}
\end{eqnarray}
After solving all the equations, it can be obtained that
$d_n^M=f_n^M=T^{TM}=0$, indicating that no field exists in the cloak
layer as well as the outside space. Meanwhile we can get that
$R^{TM}=-\zeta_n(k_1R_1)/\psi_n(k_1R_1)$, which is important for
later use. In addition, it follows that in the limit
$\delta\rightarrow0$, $f_n^{M}\chi_n^{'}(k_t\delta)$ is nonzero.
Obviously, the value of this product in Eq.(\ref{eq:E_R1}) is
nonzero only at the inner boundary, bringing out the discontinuity
of the tangential $E$ field across the inner boundary. This result
is very interesting. Since both $\mu$ and $\epsilon$ are finite
everywhere and no conductive media exist, it is unreasonable to
include surface current to support this discontinuity, as the common
boundary conditions do. It should be noted that due to the same
reason the displacement surface currents introduced in the
cylindrical cloak \cite{baile, greenleaf_OE} are not applicable in
the present case.

\begin{figure}
\includegraphics[width=0.6\columnwidth,draft=false]{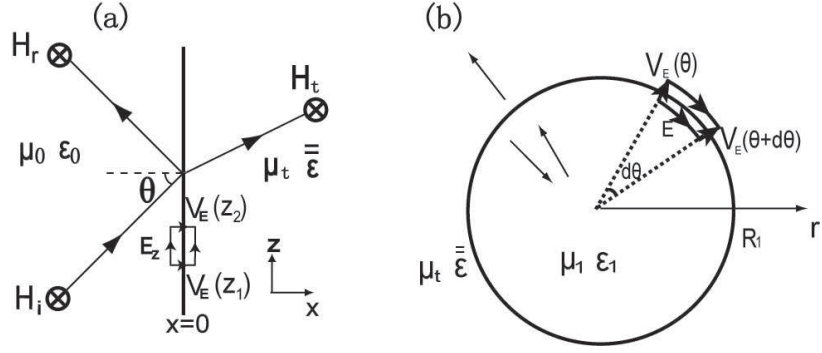}
\caption{\label{fig:half_space} Reflection and transmission of (a) a
TM wave in Cartesian coordinate system incident from free space onto
an uniaxial medium whose $\epsilon_x$ goes to zero; and (b) an
outgoing TM wave in spherical coordinate system from the homogeneous
and isotropic hole through a homogeneous uniaxial background medium
whose radial permittivity goes to zero.}
\end{figure}

In order to understand this discontinuity, let us first consider a
similar case in Cartesian coordinates as shown in
Fig.~\ref{fig:half_space}(a), where a TM wave
$\bar{H}_i=\hat{y}e^{ik_{ix}x+ik_zz}$ is obliquely incident
($k_{z}\neq0$) from free space onto an uniaxial medium with
permittivity tensor $\epsilon_x\hat x\hat x+\epsilon_t\hat y\hat
y+\epsilon_t\hat z\hat z$ and permeability $\mu_t$. The dispersion
relation in this medium is
$k_{tx}^2/(\omega^2\epsilon_t\mu_t)+k_z^2/(\omega^2\epsilon_x\mu_t)=1$.
Thus, when $\epsilon_x$ is very small, $k_{tx}$ becomes imaginary
and the transmitted wave becomes evanescent. In the limit
$\epsilon_x\rightarrow 0$, it can be found that the transmitted wave
is so strongly evanescent that no fields exist in the region $x>0$.
More interestingly, in the limit $\epsilon_x\rightarrow0$, the
integration $\int_0^\infty E_{tx}dx$ has a finite value $-2i\eta_0
\cos \theta e^{i k_z z}/k_z$. In other words, $E_{tx}$ is compressed
on the interface like a delta function. This special finite and
nonzero value can be named as an electric surface voltage $V_E$.
When a free charge $q$ moves to the interface, this voltage will
push it to the other side and transfer energy $qV_E$ to it. This
voltage is not caused by conductive charges but an infinite
polarization of the material on the interface, i.e. it corresponds
to a distribution of polarized dipole moments on the interface. In
addition, the tangential electric field at the left side of the
interface is $E_{iz}+E_{rz}=-2\eta_0 \cos \theta e^{ik_z z}$ while
that at the right side is zero, meaning the tangential $E$ field is
discontinuous across the interface. However, since
$(E_{iz}+E_{rz})\triangle z + V_E(z_2)-V_E(z_1)=0$, as shown in
Fig.~\ref{fig:half_space}(a), Faraday's law still holds on this
interface. Clearly, using this uniaxial material, which is the same
with the inner boundary of the cloak, $E_x$ becomes a delta function
and forms the electric surface voltage which supports the
discontinuity of the tangential $E$ field. Meanwhile, the reflection
coefficient becomes -1, meaning that this medium behaves like a PMC
by means of controlling the medium's electric response.

\begin{figure}
\includegraphics[width=0.7\columnwidth,draft=false]{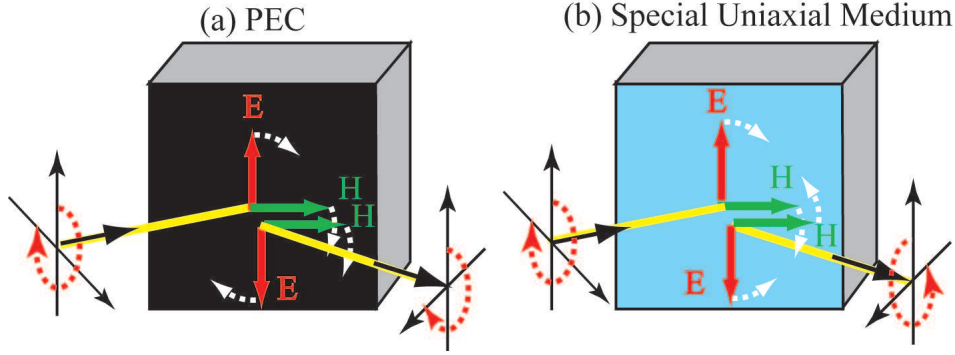}
\caption{\label{fig:reflection}(Color online) Reflection  of a
right-handed circularly polarized wave from (a) a PEC and (b) a
medium with normal permittivity and permeability go to zero
simultaneously. The dotted arrows represent the direction of
rotation.}
\end{figure}

Similarly, the reflection coefficient for a TE wave is also -1 if
$\mu_x$ goes to zero. Thus, the special uniaxial medium whose
$\epsilon_x$ and $\mu_x$ go to zero simultaneously behaves like a
PMC for TM waves due to electric surface voltages, and a PEC for TE
waves due to magnetic surface voltages. This leads to another
interesting aspect of the reflection. First, in the sense that there
is a complete reflection, the interface behaves like a mirror.
Second, this special mirror not only keeps the polarization but also
the phase information of the reflected waves. For example, for a
right-handed circularly polarized wave incident onto this boundary,
the reflected wave retains its handness, but for a mere PEC or PMC
boundary, the reflected wave becomes left-handed, as shown in
Fig.~\ref{fig:reflection}. This property is similar to the SHS
boundary used in radar and microwave engineering \cite{lindell}. But
for a SHS with its conducting vector fixed, if the incident plane
changes, the phase of the reflected wave also changes. However, the
phase of the reflected wave in Fig.~\ref{fig:reflection}(b) is
independent on the incident plane, meaning it only depends on the
optical path the wave travels. So, this mirror behaves the same in
any plane of incidence, and the information of a source including
the polarization and phase is entirely retained in the reflected
wave.

From the above discussion in Cartesian coordinate system, we see
that the surface voltages are introduced by the zero permittivity
and permeability in the normal direction of the interface, which
contribute to the discontinuity of the tangential electromagnetic
fields across the boundary. This is also true for a spherical
interface in spherical coordinate system. For example, as shown in
Fig.~\ref{fig:half_space}(b), a sphere with permittivity
$\epsilon_1$ and permeability $\mu_1$ is embedded in the homogeneous
background medium with permittivity tensor
$\bar{\bar{\epsilon}}=\epsilon_r
\hat{r}\hat{r}+\epsilon_t\hat{\theta}\hat{\theta}+\epsilon_t\hat{\phi}\hat{\phi}$
and permeability $\mu_t$. Similar to the case in
Fig.~\ref{fig:half_space}(a), for TM waves, in the limit
$\epsilon_r\rightarrow 0$, no fields exist in the region $r>R_1$,
but  the electric surface voltage $V_E$ is induced at the boundary.
Since
$E_{\theta}^{int}R_1d\theta+V_E(\theta+d\theta)-V_E(\theta)=0$,
Faraday's law still holds across the boundary. The similar result
can be obtained for $E_{\phi}$ component. Outgoing TE waves have
similar derivation when $\mu_r$ goes to zero. Therefore the
condition where the material at the inner boundary of a spherical
cloak has radial permittivity and permeability of zero is sufficient
for total reflection of all waves back by inducing surface voltages,
no matter whether the outside medium satisfies the relation of
constitutive parameters proposed in Ref.~\cite{pendry} or not.
Mathematical treatment in time domain in Ref. \cite{weder} has also
gotten the similar result of complete reflection.

\begin{figure}
\includegraphics[width=0.6\columnwidth,draft=false]{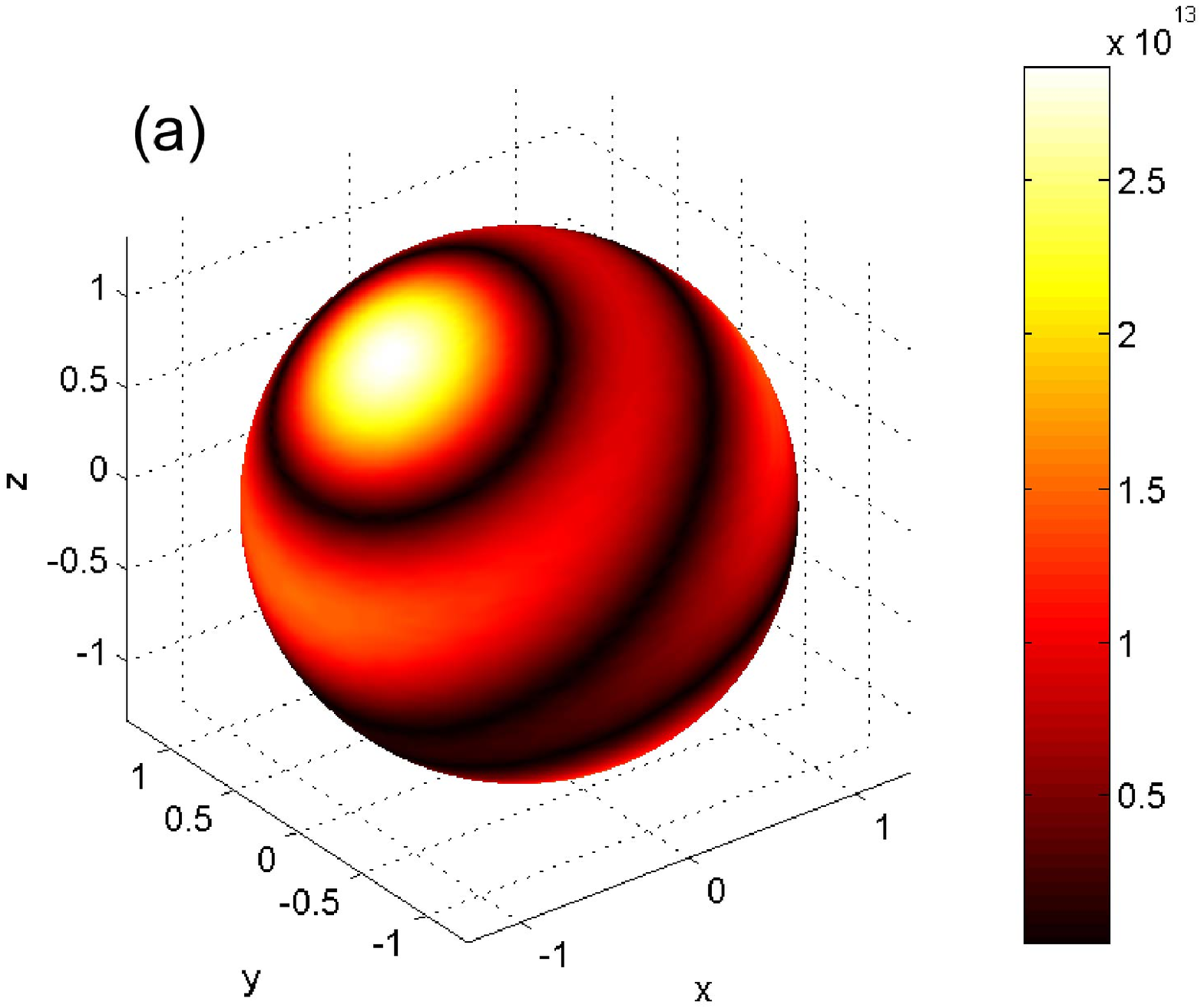} 
\includegraphics[width=0.6\columnwidth,draft=false]{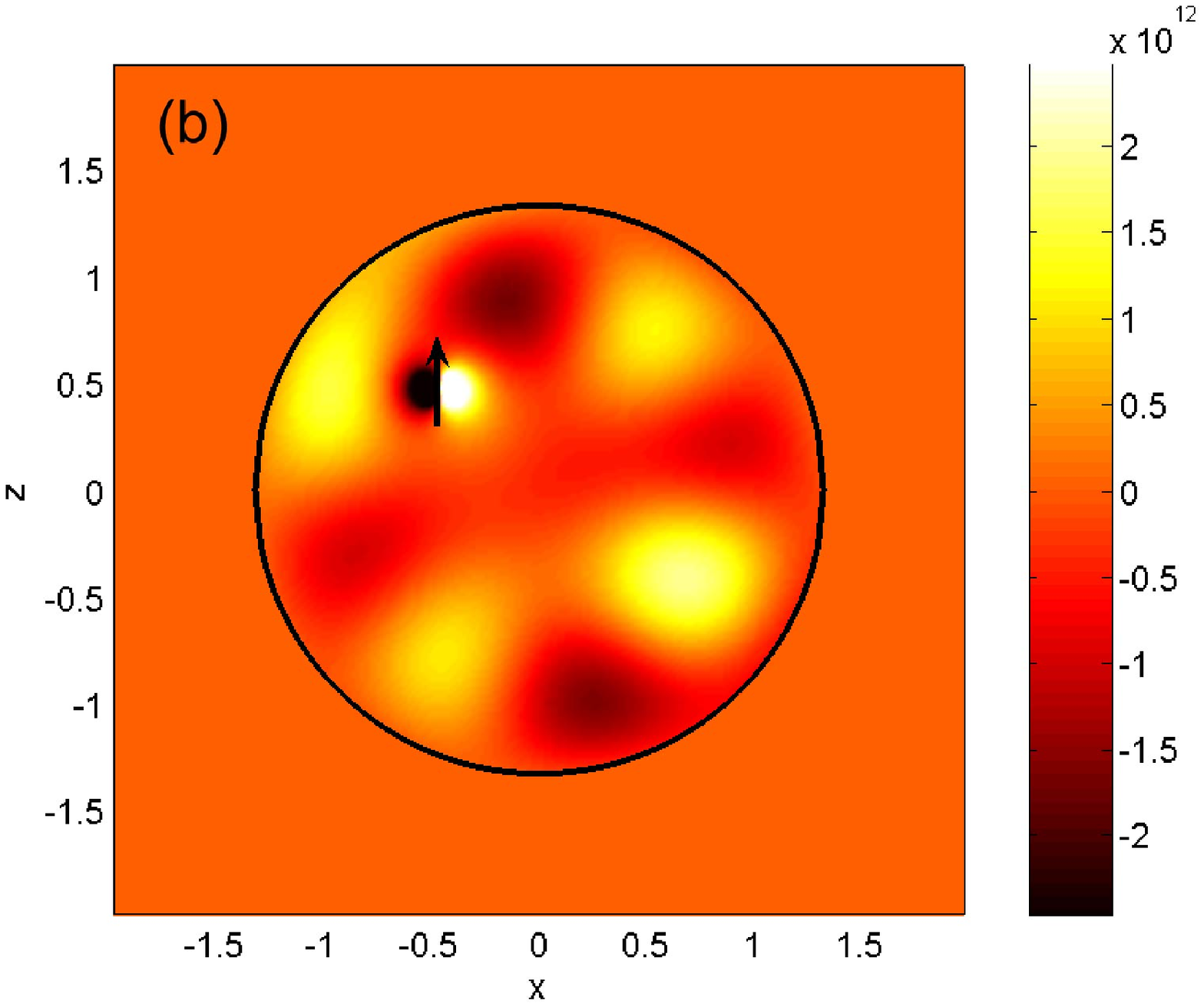} 
\caption{\label{fig:VEH}(Color online) Distributions of (a)the
amplitude of $V_E$ at the inner boundary and (b) $H_y$ in the
concealed region $r<R_1$ due to an unit electric dipole oriented
in $z$ direction and located at ($R_1/2$, $\pi/4$, $\pi$)
indicated by a small arrow in (b). $R_1=1.33\lambda_1$. }
\end{figure}

Based on the above discussion,  the electric and magnetic surface
voltages at the inner boundary as well as the field distribution
inside the concealed region due to an electric dipole $\bar{p}$
located at an arbitrary position ($r^{'}$,$\theta^{'}$,$\phi^{'}$),
where $r^{'}<R_1$, can be  derived. By expanding the wave from the
dipole into spherical waves, the corresponding scalar potentials
$\Psi_{TM}^{i}$ and $\Psi_{TE}^{i}$ for the incident waves can be
obtained. Since it is known that the reflection coefficient for both
TE and TM waves is $-\zeta_n(k_1R_1)/\psi_n(k_1R_1)$,  the scalar
potentials of reflected waves, $\Psi_{TE}^{r}$ and $\Psi_{TM}^{r}$,
can be easily obtained. Consequently, the induced electric and
magnetic surface voltages at the inner boundary of the spherical
cloak can be calculated as follows
\begin{eqnarray}
V_E=\int_{R_1^-}^{R_1^+}E_r
dr=\frac{-i}{\omega\mu_1\epsilon_1}\frac{\partial}{\partial
r}(\Psi_{TM}^{i}+\Psi_{TM}^{r})|_{r=R_1^-}\:,\label{eq:VE}
\\
V_H=\int_{R_1^-}^{R_1^+}H_r
dr=\frac{-i}{\omega\mu_1\epsilon_1}\frac{\partial}{\partial
r}(\Psi_{TE}^{i}+\Psi_{TE}^{r})|_{r=R_1^-}\:.\label{eq:VH}
\end{eqnarray}

Figure \ref{fig:VEH} plots the amplitude of $V_E$ at the inner
boundary of a spherical cloak and the field $H_y$ inside the
concealed region in the $x$-$z$ plane, due to an electric dipole
pointing in $z$ direction and located at ($R_1/2$, $\pi/4$, $\pi$).
Firstly, it is seen that surface voltages distributions are not
uniform on the surface. But for an outside observer, the dipole is
invisible since no wave propagates outside. Secondly, the field
inside exists in the form of standing waves. Figure \ref{fig:VEH}(b)
shows the field at the moment that the magnetic field reaches
maximum. After a quarter of cycle, the magnetic field becomes zero
while the electric field reaches maximum. Since $E$ and $H$ are
always out of phase, the time-averaged Poynting power is zero
anywhere, meaning no time-averaged power flowing inside. In other
words, the energy radiated from the dipole at this moment will be
returned to the dipole the next moment. Thus the total energy inside
will not blow up.

It can be calculated from Eqs.(\ref{eq:VE}) and (\ref{eq:VH}) that,
in the presence of an electric dipole inside, when $\omega$
decreases to zero, $V_H$ becomes zero while $V_E$ survives.
Similarly, if a static magnetic dipole is inside instead of a static
electric dipole, $V_E$ vanishes while $V_H$ survives. The cloak for
the static magnetic field can be realized artificially \cite{wood}.
Since there is no magnetic charge in nature, this magnetic surface
voltage induced by a static magnetic dipole must exist in the form
of its equivalent electric surface current. The inner boundary in
Ref.~\cite{wood} is made of superconductor which makes this surface
current realizable.

Furthermore, the value of surface voltages can relate to another
parameter directly. In derivation of the scalar potential in
Ref.~\cite{hongsheng}, the condition $\frac{\partial}{\partial
r}\Psi_{TM} = i\omega \epsilon \mu \varphi_e$\cite{balanis}, where
$\varphi_e$ represents the auxiliary electric scalar potential, is
applied. It is interesting to note that $V_E =
\varphi_e|_{r=R_1^-}$. Similarly, $V_H=\varphi_m|_{r=R_1^-}$. Thus
these auxiliary scalar potentials, $\varphi_e$ and $\varphi_m$,
which were introduced originally as mathematical tools, have direct
physical counterparts at the inner boundary of the cloak, i.e.
surface voltages in this case.

In conclusion, the exact electromagnetic field solution to the
Maxwell equations for a spherical cloak with an active source inside
the concealed region is established. It is shown that, an infinite
polarization of the material at the inner boundary of the cloak will
induce the electric and magnetic surface voltages, which prevent all
waves from going out. These peculiar surface voltages are rare in
nature, but they do not violate the Maxwell equations. The handness
of the polarization and phase information of the waves reflected
from the inner boundary of the cloak are unchanged. Finally, these
surface voltages due to an electric dipole inside the concealed
region are found to be exactly equal to the auxiliary scalar
potentials at the inner boundary of the cloak which gain physical
counterparts in this case.

This work is supported by the ONR under Contract N00014-01-1-0713,
the Chinese NSF under Grant 60531020, and China PSF under Grant
20060390331.

\bibliographystyle{apsrev}

\begin{thebibliography}{18}
\expandafter\ifx\csname
natexlab\endcsname\relax\def\natexlab#1{#1}\fi
\expandafter\ifx\csname bibnamefont\endcsname\relax
  \def\bibnamefont#1{#1}\fi
\expandafter\ifx\csname bibfnamefont\endcsname\relax
  \def\bibfnamefont#1{#1}\fi
\expandafter\ifx\csname citenamefont\endcsname\relax
  \def\citenamefont#1{#1}\fi
\expandafter\ifx\csname url\endcsname\relax
  \def\url#1{\texttt{#1}}\fi
\expandafter\ifx\csname urlprefix\endcsname\relax\def\urlprefix{URL
}\fi \providecommand{\bibinfo}[2]{#2}
\providecommand{\eprint}[2][]{\url{#2}}

\bibitem[{\citenamefont{Pendry et~al.}(2006)\citenamefont{Pendry, Schurig, and
  Smith}}]{pendry}
\bibinfo{author}{\bibfnamefont{J.~B.} \bibnamefont{Pendry}},
  \bibinfo{author}{\bibfnamefont{D.}~\bibnamefont{Schurig}}, \bibnamefont{and}
  \bibinfo{author}{\bibfnamefont{D.~R.} \bibnamefont{Smith}},
  \bibinfo{journal}{Science} \textbf{\bibinfo{volume}{312}},
  \bibinfo{pages}{1780} (\bibinfo{year}{2006}).

\bibitem[{\citenamefont{Leonhardt}(2006)}]{leonhardt}
\bibinfo{author}{\bibfnamefont{U.}~\bibnamefont{Leonhardt}},
  \bibinfo{journal}{Science} \textbf{\bibinfo{volume}{312}},
  \bibinfo{pages}{1777} (\bibinfo{year}{2006}).

\bibitem[{\citenamefont{Alu and Engheta}(2005)}]{alu}
\bibinfo{author}{\bibfnamefont{A.}~\bibnamefont{Alu}} \bibnamefont{and}
  \bibinfo{author}{\bibfnamefont{N.}~\bibnamefont{Engheta}},
  \bibinfo{journal}{Phys. Rev. E} \textbf{\bibinfo{volume}{72}},
  \bibinfo{pages}{016623} (\bibinfo{year}{2005}).

\bibitem[{\citenamefont{Milton and Nicorovici}(2006)}]{milton}
\bibinfo{author}{\bibfnamefont{G.~W.} \bibnamefont{Milton}} \bibnamefont{and}
  \bibinfo{author}{\bibfnamefont{N.-A.~P.} \bibnamefont{Nicorovici}},
  \bibinfo{journal}{Proc. R. Soc. A} \textbf{\bibinfo{volume}{462}},
  \bibinfo{pages}{3027} (\bibinfo{year}{2006}).


\bibitem[{\citenamefont{Sihvola}(2006)}]{sihvola}
\bibinfo{author}{\bibfnamefont{A.}~\bibnamefont{Sihvola}},
  \bibinfo{journal}{Prog. Electromagn. Res.}
  \textbf{\bibinfo{volume}{pier-66}}, \bibinfo{pages}{191}
  (\bibinfo{year}{2006}).

\bibitem[{\citenamefont{Greenleaf et~al.}(2007)\citenamefont{Greenleaf,
  Lassas, and Uhlmann}}]{greenleaf_static}
\bibinfo{author}{\bibfnamefont{A.}~\bibnamefont{Greenleaf}},
  \bibinfo{author}{\bibfnamefont{M.}~\bibnamefont{Lassas}}, \bibnamefont{and}
  \bibinfo{author}{\bibfnamefont{G.}~\bibnamefont{Uhlmann}},
  \bibinfo{journal}{ Math. Res. Lett.} \textbf{\bibinfo{volume}{10}},
  \bibinfo{pages}{685}(\bibinfo{year}{2003}).

\bibitem[{\citenamefont{Sheng}(2006)}]{sheng}
\bibinfo{author}{\bibfnamefont{P.}~\bibnamefont{Sheng}},
  \bibinfo{journal}{Science} \textbf{\bibinfo{volume}{313}},
  \bibinfo{pages}{1399} (\bibinfo{year}{2006}).

\bibitem[{\citenamefont{Schurig
  et~al.}(2006{\natexlab{a}})\citenamefont{Schurig, Pendry, and
  Smith}}]{schurig_ray}
\bibinfo{author}{\bibfnamefont{D.}~\bibnamefont{Schurig}},
  \bibinfo{author}{\bibfnamefont{J.~B.} \bibnamefont{Pendry}},
  \bibnamefont{and} \bibinfo{author}{\bibfnamefont{D.~R.} \bibnamefont{Smith}},
  \bibinfo{journal}{Opt. Express} \textbf{\bibinfo{volume}{14}},
  \bibinfo{pages}{9794} (\bibinfo{year}{2006}{\natexlab{a}}).

\bibitem[{\citenamefont{Cummer et~al.}(2006)\citenamefont{Cummer, Popa,
  Schurig, Smith, and Pendry}}]{cummer}
\bibinfo{author}{\bibfnamefont{S.~A.} \bibnamefont{Cummer}},
  \bibinfo{author}{\bibfnamefont{B.-I.} \bibnamefont{Popa}},
  \bibinfo{author}{\bibfnamefont{D.}~\bibnamefont{Schurig}},
  \bibinfo{author}{\bibfnamefont{D.~R.} \bibnamefont{Smith}}, \bibnamefont{and}
  \bibinfo{author}{\bibfnamefont{J.~B.} \bibnamefont{Pendry}},
  \bibinfo{journal}{Phys. Rev. E} \textbf{\bibinfo{volume}{74}},
  \bibinfo{pages}{036621} (\bibinfo{year}{2006}).

\bibitem[{\citenamefont{Zolla et~al.}(2007)\citenamefont{Zolla, Guenneau,
  Nicolet, and Pendry}}]{zolla}
\bibinfo{author}{\bibfnamefont{F.}~\bibnamefont{Zolla}},
  \bibinfo{author}{\bibfnamefont{S.}~\bibnamefont{Guenneau}},
  \bibinfo{author}{\bibfnamefont{A.}~\bibnamefont{Nicolet}}, \bibnamefont{and}
  \bibinfo{author}{\bibfnamefont{J.~B.} \bibnamefont{Pendry}},
  \bibinfo{journal}{Opt. Lett.} \textbf{\bibinfo{volume}{32}},
  \bibinfo{pages}{1069} (\bibinfo{year}{2007}).

\bibitem[{\citenamefont{Chen et~al.}(2007)\citenamefont{Chen, Wu, Zhang, and
  Kong}}]{hongsheng}
\bibinfo{author}{\bibfnamefont{H.}~\bibnamefont{Chen}},
  \bibinfo{author}{\bibfnamefont{B.~I.} \bibnamefont{Wu}},
  \bibinfo{author}{\bibfnamefont{B.}~\bibnamefont{Zhang}}, \bibnamefont{and}
  \bibinfo{author}{\bibfnamefont{J.~A.} \bibnamefont{Kong}},
  \bibinfo{journal}{Phys. Rev. Lett.} \textbf{\bibinfo{volume}{99}},
  \bibinfo{pages}{063903} (\bibinfo{year}{2007}).

\bibitem[{\citenamefont{Zhang et~al.}(2007)\citenamefont{Zhang, Chen, Wu, Luo,
  Ran, and Kong}}]{baile}
\bibinfo{author}{\bibfnamefont{B.}~\bibnamefont{Zhang}},
  \bibinfo{author}{\bibfnamefont{H.}~\bibnamefont{Chen}},
  \bibinfo{author}{\bibfnamefont{B.~I.} \bibnamefont{Wu}},
  \bibinfo{author}{\bibfnamefont{Y.}~\bibnamefont{Luo}},
  \bibinfo{author}{\bibfnamefont{L.}~\bibnamefont{Ran}}, \bibnamefont{and}
  \bibinfo{author}{\bibfnamefont{J.~A.} \bibnamefont{Kong}},
  \bibinfo{journal}{Phys. Rev. B} \textbf{\bibinfo{volume}{76}},
  \bibinfo{pages}{121101(R)} (\bibinfo{year}{2007}).

\bibitem[{\citenamefont{Schurig
  et~al.}(2006{\natexlab{b}})\citenamefont{Schurig, Mock, Justice, Cummer,
  Pendry, Starr, and Smith}}]{schurig}
\bibinfo{author}{\bibfnamefont{D.}~\bibnamefont{Schurig}},
  \bibinfo{author}{\bibfnamefont{J.~J.} \bibnamefont{Mock}},
  \bibinfo{author}{\bibfnamefont{B.~J.} \bibnamefont{Justice}},
  \bibinfo{author}{\bibfnamefont{S.~A.} \bibnamefont{Cummer}},
  \bibinfo{author}{\bibfnamefont{J.~B.} \bibnamefont{Pendry}},
  \bibinfo{author}{\bibfnamefont{A.~F.} \bibnamefont{Starr}}, \bibnamefont{and}
  \bibinfo{author}{\bibfnamefont{D.~R.} \bibnamefont{Smith}},
  \bibinfo{journal}{Science} \textbf{\bibinfo{volume}{314}},
  \bibinfo{pages}{977} (\bibinfo{year}{2006}{\natexlab{b}}).

\bibitem[{\citenamefont{Greenleaf et~al.}(2007)\citenamefont{Greenleaf,
  Kurylev, Lassas, and Uhlmann}}]{greenleaf}
\bibinfo{author}{\bibfnamefont{A.}~\bibnamefont{Greenleaf}},
  \bibinfo{author}{\bibfnamefont{Y.}~\bibnamefont{Kurylev}},
  \bibinfo{author}{\bibfnamefont{M.}~\bibnamefont{Lassas}}, \bibnamefont{and}
  \bibinfo{author}{\bibfnamefont{G.}~\bibnamefont{Uhlmann}},
  \bibinfo{journal}{Commun. Math. Phys.} \textbf{\bibinfo{volume}{275}},
  \bibinfo{pages}{749}
  (\bibinfo{year}{2007}).


\bibitem[{\citenamefont{Lindell and Puska}(1996)}]{lindell}
\bibinfo{author}{\bibfnamefont{I.~V.} \bibnamefont{Lindell}} \bibnamefont{and}
  \bibinfo{author}{\bibfnamefont{P.}~\bibnamefont{Puska}},
  \bibinfo{journal}{IEE Proc.-Microw. Antennas Propag.}
  \textbf{\bibinfo{volume}{143}}, \bibinfo{pages}{417} (\bibinfo{year}{1996}).


\bibitem[{\citenamefont{Balanis}()}]{balanis}
\bibinfo{author}{\bibfnamefont{C.~A.} \bibnamefont{Balanis}},
  \eprint{\textit{Advanced Engineering Electromagnetics}, John Wiley\&Sons,
  1989}.


\bibitem[{\citenamefont{Chew}()}]{weng}
\bibinfo{author}{\bibfnamefont{W.~C.} \bibnamefont{Chew}},
  \eprint{\textit{Waves and Fields in inhomogeneous Media}, 2nd ed, IEEE Press,
  1995}.

\bibitem[{\citenamefont{Greenleaf et~al.}(2007)\citenamefont{Greenleaf,
  Kurylev, Lassas, and Uhlmann}}]{greenleaf_OE}
\bibinfo{author}{\bibfnamefont{A.}~\bibnamefont{Greenleaf}},
  \bibinfo{author}{\bibfnamefont{Y.}~\bibnamefont{Kurylev}},
  \bibinfo{author}{\bibfnamefont{M.}~\bibnamefont{Lassas}}, \bibnamefont{and}
  \bibinfo{author}{\bibfnamefont{G.}~\bibnamefont{Uhlmann}},
  \bibinfo{journal}{Opt. Express} \textbf{\bibinfo{volume}{15}},
  \bibinfo{pages}{12717}
  (\bibinfo{year}{2007}).



\bibitem[{\citenamefont{Weder et~al.}(2007)\citenamefont{Weder}}]{weder}
\bibinfo{author}{\bibfnamefont{R.}~\bibnamefont{Weder}},
  \bibinfo{journal}{arXiv:0704.0248v3}
  (\bibinfo{year}{2007}).



\bibitem[{\citenamefont{Wood and Pendry}(2007)}]{wood}
\bibinfo{author}{\bibfnamefont{B.}~\bibnamefont{Wood}} \bibnamefont{and}
  \bibinfo{author}{\bibfnamefont{J.~B.} \bibnamefont{Pendry}},
  \bibinfo{journal}{J. Phys.: Condens. Matter} \textbf{\bibinfo{volume}{19}},
  \bibinfo{pages}{076208} (\bibinfo{year}{2007}).

\end{thebibliography}

\end{document}